\documentclass[11pt]{article}
\usepackage{epsfig} 
\newcommand{\sect}[1]{ \section{#1} \setcounter{equation}{0} }

\newcommand{\Dslash}{D \! \! \! \! /}

\newcommand{\half}{\mbox{\small{$\frac{1}{2}$}}} 
 
\newcommand{\Nf}{N_{\!f}} 
\newcommand{\MSbar}{\overline{\mbox{MS}}}

\begin{document}
\title{Two loop $\MSbar$ gluon pole mass from the LCO formalism} 
\author{J.A. Gracey, \\ Theoretical Physics Division, \\ 
Department of Mathematical Sciences, \\ University of Liverpool, \\ P.O. Box 
147, \\ Liverpool, \\ L69 3BX, \\ United Kingdom.} 
\date{} 
\maketitle 
\vspace{5cm} 
\noindent 
{\bf Abstract.} We compute the pole mass of the gluon in QCD from the local 
composite operator formalism at two loops in the $\MSbar$ renormalization 
scheme. For Yang-Mills theory an estimate of the mass at two loops is 
$2.13\Lambda_{\mbox{\footnotesize{$\MSbar$}}}$.  

\vspace{-16cm}
\hspace{13.5cm}
{\bf LTH 642}

\newpage

\sect{Introduction.}
There has been recent interest in understanding the role the dimension two
gauge invariant gluon mass operator plays in the vacuum structure of Yang-Mills
theory and QCD. For instance, see \cite{1,2,3,4,5,6,7,8,9,10,11,12,13} and
references therein. As was pointed out in \cite{14,15,16,17} the perturbative 
vacuum one ordinarily uses for high energy computations is not stable and it 
has been suggested that in the true vacuum, various operators condense 
developing non-zero vacuum expectation values. Whilst the operator product 
expansion and QCD sum rules are usually centred on the gauge invariant 
operators $\left(G^a_{\mu\nu}\right)^2$ and $\bar{\psi}\psi$, where 
$G^a_{\mu\nu}$ is the gluon field strength and $\psi$ is the quark field, 
recent work considered the lower dimension operator $\half A^{a\,2}_\mu$ in the
Landau gauge, \cite{3,8,9,10,11,12,13}, and various generalizations of it. For
instance, there one can construct a non-local gauge invariant dimension two 
operator which truncates to $\half A^{a\,2}_\mu$ in the Landau gauge. An
example of the role such an operator can play in the infrared structure has 
been examined in \cite{18} where a massive gauge invariant QCD Lagrangian was 
studied in connection with vortex solutions. As $\half A_\mu^{a\,2}$ has the 
dimensions of a mass operator, it has been the subject of investigating the 
issue of whether the gluon can develop a mass dynamically. Indeed in the early 
work of Curci and Ferrari, \cite{19}, an extension of this operator, which was 
on-shell BRST invariant, was included in the usual QCD Lagrangian. More 
recently another approach, known as the local composite operator (LCO) method, 
has been developed which avoids the ad hoc inclusion of a gluon mass term 
\cite{8,9,10}. Instead the QCD Lagrangian is modified to introduce an extra 
scalar field, $\sigma$, coupled to $\half A^{a\,2}_\mu$ and with this one can 
compute the effective potential of the scalar field. It transpires that due to 
the development of a non-zero vacuum expectation value for $\sigma$ the gluon 
gains a non-zero mass in a vacuum which does not correspond to the (unstable) 
perturbative one. By contrast, applying the same formalism to QED, \cite{10}, 
the perturbative vaccum is stable and whilst there is another extremum where 
$\sigma$ has a non-zero vacuum expectation value, it corresponds to an unstable
point.

Within this formalism one can estimate the size of an effective gluon mass at 
one and two loops. In Yang-Mills theory it is of the order of 
$2\Lambda_{\mbox{\footnotesize{$\MSbar$}}}$ and is stable to the higher order
corrections. Whilst this is roughly consistent with other estimates of a gluon
mass from a wide range of methods (which are succinctly summarized in Table 15
of Field's article, \cite{20}) the LCO estimates suffer several shortcomings.
One of these is that the effective gluon mass used in \cite{8} was the tree
object and whilst the effective potential does have the quantum corrections no 
account of the dressings of the tree quantity were included. Further, whilst 
all the gluon estimates are of a similar range, it is not clear to what extent 
the same mass quantity is being measured. For instance, in the quark sector of 
QCD the quark masses are all measured and compared to the same benchmark, which
is the running mass at the scale $2$GeV. This is irrespective of whether the 
pole mass of the quark was determined or, say, the running mass at another 
scale prior to using the (four loop) quark mass anomalous dimension to run the 
mass to the standard reference scale. For the same problem for a gluon mass, 
the anomalous dimension of the $\half A^{a\,2}_\mu$ gluon mass operator in the 
Landau gauge is now available at four loops, \cite{21}, extending the two,
\cite{22}, and three loop, \cite{23}, results which are all in the $\MSbar$
scheme. Remarkably the operator anomalous dimension is the sum of the gluon and
ghost anomalous dimensions in the Landau gauge, \cite{23,24}. To complete the 
analysis for any future gluon mass computations one requires, for instance, the
relation between the running and pole mass of the gluon. This was initially 
addressed for the LCO formalism in \cite{25} where the one loop relation 
between these quantities was given in the $\MSbar$ scheme where the computation
extended the one loop calculation of \cite{26} for the Curci-Ferrari Lagrangian
(with $\Nf$~$=$~$0$) itself rather than the LCO one. Moreover, an estimate of 
the pole mass was provided, \cite{25}, by converting the effective potential of
the classical gluon mass of the LCO Lagrangian into a potential for the gluon 
pole mass. Remarkably upon extremization and solution of the resultant 
equation, a mass estimate emerged in QCD which was {\em independent} of the 
renormalization scale. For Yang-Mills theory this corresponded to 
$2.10\Lambda_{\mbox{\footnotesize{$\MSbar$}}}$, \cite{25}, though the values 
for $\Nf$~$=$~$2$ and $3$ were significantly lower. Given the interest in gluon
masses and the absence of relations between the various mass quantities, it is
the purpose of this article to extend \cite{25} to two loops by computing the 
gluon pole mass in the LCO formalism in the $\MSbar$ scheme in the Landau
gauge. As a by-product we will be to deduce the same quantity in the 
Curci-Ferrari Lagrangian to extend the Yang-Mills result of \cite{26}. Another 
motivation for such an analysis, aside from determining whether the two loop 
corrections significantly alter the one loop estimates, is to ascertain whether
the two loop result using the minimization criterion of \cite{25} retains the 
one loop renormalization scale independence. Given the fact that by analogy, 
except for the $\beta$-function, the $\MSbar$ anomalous dimensions are 
renormalization scheme independent at only {\em one} loop, it would be 
surprising if a two loop pole mass estimate remained renormalization scale 
independent. 

The article is organised as follows. In section $2$, we review the background
to the problem including the necessary points of the LCO formalism before
discussing the construction of the two loop gluon pole mass in section $3$.
Equipped with this we perform the analysis to produce a mass estimate in 
section $4$ before concluding with various estimates in section $5$. 

\sect{LCO formalism.} 
We begin by reviewing the key ingredients of the LCO formalism, \cite{8}, we
will require. First, we define the QCD Lagrangian in an arbitrary covariant 
gauge as 
\begin{equation} 
L^{\mbox{\footnotesize{QCD}}} ~=~ -~ \frac{1}{4} G_{\mu\nu}^a 
G^{a \, \mu\nu} ~-~ \frac{1}{2\alpha} (\partial^\mu A^a_\mu)^2 ~-~ 
\bar{c}^a \partial^\mu D_\mu c^a ~+~ i \bar{\psi}^{iI} \Dslash \psi^{iI} 
\end{equation} 
where $\alpha$ is the gauge fixing parameter, $A^a_\mu$ is the gluon field,
$c^a$ and $\bar{c}^a$ are the ghost and anti-ghost fields, $\psi^{iI}$ is the
quark field and the various indices range over $1$~$\leq$~$a$~$\leq$~$N_A$, 
$1$~$\leq$~$I$~$\leq$~$N_F$ and $1$~$\leq$~$i$~$\leq$~$\Nf$ where $N_F$ and 
$N_A$ are the dimensions of the fundamental and adjoint representations 
respectively and $\Nf$ is the number of quarks. To construct the LCO Lagrangian
from $L^{\mbox{\footnotesize{QCD}}}$ we introduce the path integral $W[J]$ 
defined by, 
\cite{8}, 
\begin{equation}
e^{-W[J]} ~=~ \int {\cal D} A^\mu_{\mbox{\footnotesize{o}}} 
{\cal D} \psi_{\mbox{\footnotesize{o}}} {\cal D} 
\bar{\psi}_{\mbox{\footnotesize{o}}} {\cal D} c_{\mbox{\footnotesize{o}}} 
{\cal D} \bar{c}_{\mbox{\footnotesize{o}}} \,
\exp \left[ \int d^d x \left( L_{\mbox{\footnotesize{o}}} ~-~ \half 
J_{\mbox{\footnotesize{o}}} A_{{\mbox{\footnotesize{o}}} \, \mu}^{a \, 2} ~+~ 
\half \xi_{\mbox{\footnotesize{o}}} J_{\mbox{\footnotesize{o}}}^2 
\right) \right]
\end{equation}  
where $J$ is the source coupled to the local composite operator 
$\half A^{a\,2}_\mu$ in the Landau gauge and the subscript 
${}_{\mbox{\footnotesize{o}}}$ denotes bare quantities. To retain 
renormalizability of the action including the source as well as a homogeneous 
renormalization group equation an additional term has been introduced. For 
instance, the term quadratic in $J$ is necessary since the vacuum energy is 
divergent as can easily be seen by power counting. This term is coupled in via 
a parameter $\xi$ and its associated counterterm $\delta \xi$ is included 
when the action is converted to renormalized parameters giving 
\begin{equation}
e^{-W[J]} ~=~ \int {\cal D} A_\mu {\cal D} \psi {\cal D} \bar{\psi} {\cal D} c 
{\cal D} \bar{c} \, \exp \left[ \int d^d x \left( L ~-~ \half Z_m J 
A_\mu^{a \, 2} ~+~ \half ( \xi + \delta \xi ) J^2 \right) \right] 
\end{equation}  
where $Z_m$ is the gluon mass renormalization constant. It transpires, 
\cite{8}, that in the LCO formalism one can compute the explicit form of $\xi$ 
by ensuring that $W[J]$ does indeed satisfy a homogeneous renormalization group
equation. Consequently since the coupling constant, $g$, runs in such an 
equation $\xi$ is constrained to satisfy a differential equation dependent on 
the $\beta$-function and anomalous dimension of the gluon mass operator, 
$\half A^{a\,2}_\mu$. This equation can be solved in a coupling constant 
expansion. As we will require the explicit form here we note that in QCD we 
have, \cite{8,10},  
\begin{eqnarray}
\frac{1}{g^2\xi(g)} &=& \left[ \frac{( 13 C_A - 8 T_F \Nf )}{9N_A} 
\right. \nonumber \\
&& \left. +~ \left( 2685464 C_A^3 T_F \Nf - 1391845 C_A^4 
- 213408 C_A^2 C_F T_F \Nf - 1901760 C_A^2 T_F^2 \Nf^2 
\right. \right. \nonumber \\
&& \left. \left. ~~~~~ 
+~ 221184 C_A C_F T_F^2 \Nf^2 + 584192 C_A \Nf^3 T_F^3 
- 55296 C_F T_F^3 \Nf^3 
\right. \right. \nonumber \\
&& \left. \left. ~~~~~ 
-~ 65536 T_F^4 \Nf^4 \right) \frac{g^2}{5184 \pi^2 N_A (35 C_A-16 T_F \Nf) 
(19 C_A-8 T_F \Nf)} 
\right. \nonumber \\
&& \left. 
+~ \left( \left( 62228252520 C_A^6 \Nf T_F - 8324745975 C_A^7 
- 42525100800 C_A^5 C_F \Nf T_F 
\right. \right. \right. \nonumber \\
&& \left. \left. \left. ~~~~~ 
-~ 123805256256 C_A^5 \Nf^2 T_F^2 + 105262940160 C_A^4 C_F \Nf^2 T_F^2
\right. \right. \right. \nonumber \\
&& \left. \left. \left. ~~~~~ 
+~ 112398515712 C_A^4 \Nf^3 T_F^3 
- 103719518208 C_A^3 C_F \Nf^3 T_F^3 
\right. \right. \right. \nonumber \\
&& \left. \left. \left. ~~~~~ 
-~ 52888043520 C_A^3 \Nf^4 T_F^4 + 50866421760 C_A^2 C_F \Nf^4 T_F^4
\right. \right. \right. \nonumber \\
&& \left. \left. \left. ~~~~~ 
+~ 12606898176 C_A^2 \Nf^5 T_F^5 - 12419334144 C_A C_F \Nf^5 T_F^5
\right. \right. \right. \nonumber \\
&& \left. \left. \left. ~~~~~ 
-~ 1207959552 C_A \Nf^6 T_F^6 + 1207959552 C_F \Nf^6 T_F^6 \right) \zeta(3)
- 13223737800 C_A^7 
\right. \right. \nonumber \\
&& \left. \left. ~~~~~ 
+~ 5886241060 C_A^6 \Nf T_F + 52585806000 C_A^5 C_F \Nf T_F 
+ 41351916768 C_A^5 \Nf^2 T_F^2 
\right. \right. \nonumber \\
&& \left. \left. ~~~~~ 
+~ 522849600 C_A^4 C_F^2 \Nf T_F - 130596636288 C_A^4 C_F \Nf^2 T_F^2 
\right. \right. \nonumber \\
&& \left. \left. ~~~~~ 
-~ 67857620736 C_A^4 \Nf^3 T_F^3 - 1286267904 C_A^3 C_F^2 \Nf^2 T_F^2 
\right. \right. \nonumber \\
&& \left. \left. ~~~~~ 
+~ 128750638080 C_A^3 C_F \Nf^3 T_F^3 + 46700324864 C_A^3 \Nf^4 T_F^4 
\right. \right. \nonumber \\
&& \left. \left. ~~~~~ 
+~ 1180127232 C_A^2 C_F^2 \Nf^3 T_F^3 - 63001780224 C_A^2 C_F \Nf^4 T_F^4 
\right. \right. \nonumber \\
&& \left. \left. ~~~~~ 
-~ 16782753792 C_A^2 \Nf^5 T_F^5 - 475987968 C_A C_F^2 \Nf^4 T_F^4 
\right. \right. \nonumber \\
&& \left. \left. ~~~~~ 
+~ 15308685312 C_A C_F \Nf^5 T_F^5 + 3106406400 C_A \Nf^6 T_F^6 
\right. \right. \nonumber \\
&& \left. \left. ~~~~~ 
+~ 70778880 C_F^2 \Nf^5 T_F^5 - 1478492160 C_F \Nf^6 T_F^6 
\right. \right. \nonumber \\
&& \left. \left. ~~~~~ 
-~ 234881024 \Nf^7 T_F^7 \right) 
\frac{g^4}{995328 \pi^4 N_A (35 C_A-16 T_F \Nf)^2 (19 C_A-8 T_F \Nf)^2} 
\right] \nonumber \\
&& +~ O(g^6) 
\label{xieqn} 
\end{eqnarray}
where $\mbox{Tr}\left( T^a T^b \right)$~$=$~$T_F\delta^{ab}$, $T^a$ is the
group generator, $C_A$ and $C_F$ are the usual colour group Casimirs and
$\zeta(z)$ is the Riemann zeta function. Consequently one uses a 
Hubbard-Stratanovich transformation to rewrite the exponential in the path 
integral of $W[J]$. This introduces the additional scalar field $\sigma$ and 
gives a generating functional where the source $J$ now couples linearly to a 
field as opposed to a composite operator. Thus, \cite{8},  
\begin{equation}
e^{-W[J]} ~=~ \int {\cal D} A_\mu {\cal D} \psi {\cal D} \bar{\psi} {\cal D} c 
{\cal D} \bar{c} {\cal D} \sigma \, \exp \left[ \int d^d x \left( L^\sigma ~-~ 
\frac{\sigma J}{g} \right) \right] 
\end{equation}  
where $L^\sigma$ is the LCO Lagrangian and is given by, \cite{8},  
\begin{eqnarray} 
L^\sigma &=& -~ \frac{1}{4} G_{\mu\nu}^a G^{a \, \mu\nu} ~-~ \frac{1}{2\alpha} 
(\partial^\mu A^a_\mu)^2 ~-~ \bar{c}^a \partial^\mu D_\mu c^a ~+~ 
i \bar{\psi}^{iI} \Dslash \psi^{iI} \nonumber \\ 
&& -~ \frac{\sigma^2}{2g^2 \xi(g) Z_\xi} ~+~ \frac{Z_m}{2 g \xi(g) 
Z_\xi} \sigma A^a_\mu A^{a \, \mu} ~-~ \frac{Z_m^2}{8\xi(g) Z_\xi} 
\left( A^a_\mu A^{a \, \mu} \right)^2 
\label{siglag} 
\end{eqnarray} 
where the first few terms of $\xi(g)$ are given by (\ref{xieqn}).  

Using (\ref{siglag}) the effective potential $V(\sigma)$ was constructed by
standard methods at two loops in the $\MSbar$ scheme, \cite{8,9,10}. The 
divergences in the vacuum energy are removed by straightforward renormalization
since the Lagrangian, $L^\sigma$, retains the renormalizability property. As 
the two loop effective potential will be required later we recall its explicit 
form is  
\begin{eqnarray}
V(\sigma) &=& 
\frac{9N_A}{2} \lambda_1 \sigma^{\prime \, 2} \nonumber \\
&& +~ \left[ \frac{3}{64} \ln \left( \frac{g \sigma^\prime}{{\mu}^2} 
\right)
+ C_A \left(
-~ \frac{351}{8} C_F \lambda_1 \lambda_2
+ \frac{351}{16} C_F \lambda_1 \lambda_3
- \frac{249}{128} \lambda_2
+ \frac{27}{64} \lambda_3
\right)
\right. \nonumber \\
&& \left. ~~~~~ 
+~ C_A^2 \left(
-~ \frac{81}{16} \lambda_1 \lambda_2
+ \frac{81}{32} \lambda_1 \lambda_3
\right)
+ \left(
-~ \frac{13}{128}
- \frac{207}{32} C_F \lambda_2
+ \frac{117}{32} C_F \lambda_3
\right)
\right] \frac{g^2 N_A \sigma^{\prime \, 2}}{\pi^2} 
\nonumber \\
&& +~ \left[ C_A 
\left(
-~ \frac{593}{16384}
- \frac{255}{16} C_F \lambda_2
+ \frac{36649}{4096} C_F \lambda_3
- \frac{1053}{64} C_F^2 \lambda_1 \lambda_2
+ \frac{1053}{128} C_F^2 \lambda_1 \lambda_3
\right. \right. \nonumber \\
&& \left. \left. ~~~~~~~~~~~
-~ \frac{5409}{1024} C_F^2 \lambda_2^2
+ \frac{1053}{1024} C_F^2 \lambda_3^2
+ \frac{891}{8192} s_2
- \frac{1}{4096} \zeta(2)
- \frac{3}{64} \zeta(3)
\right. \right. \nonumber \\
&& \left. \left. ~~~~~~~~~~~
+~ \frac{585}{16} \zeta(3) C_F \lambda_2
- \frac{4881}{256} \zeta(3) C_F \lambda_3
\right)
\right. \nonumber \\
&& \left. ~~~~~
+~ C_A^2 \left(
-~ \frac{11583}{128} C_F \lambda_1 \lambda_2
+ \frac{11583}{256} C_F \lambda_1 \lambda_3
+ \frac{72801}{2048} C_F \lambda_2^2
+ \frac{11583}{2048} C_F \lambda_3^2
\right. \right. \nonumber \\
&& \left. \left. ~~~~~~~~~~~~~~~
+~ \frac{3159}{128} C_F^2 \lambda_1 \lambda_2^2
+ \frac{3159}{512} C_F^2 \lambda_1 \lambda_3^2
+ \frac{372015}{16384} \lambda_2
- \frac{189295}{16384} \lambda_3
\right. \right. \nonumber \\
&& \left. \left. ~~~~~~~~~~~~~~~
+~ \frac{3159}{16} \zeta(3) C_F \lambda_1 \lambda_2
- \frac{3159}{32} \zeta(3) C_F \lambda_1 \lambda_3
- \frac{1053}{16} \zeta(3) C_F \lambda_2^2
\right. \right. \nonumber \\
&& \left. \left. ~~~~~~~~~~~~~~~
-~ \frac{3159}{256} \zeta(3) C_F \lambda_3^2
- \frac{6885}{256} \zeta(3) \lambda_2
+ \frac{116115}{8192} \zeta(3) \lambda_3
\right)
\right. \nonumber \\
&& \left. ~~~~~
+~ C_A^3 \left(
\frac{34749}{256} C_F \lambda_1 \lambda_2^2
+ \frac{34749}{1024} C_F \lambda_1 \lambda_3^2
+ \frac{64071}{512} \lambda_1 \lambda_2
- \frac{64071}{1024} \lambda_1 \lambda_3
\right. \right. \nonumber \\
&& \left. \left. ~~~~~~~~~~~~~~~
-~ \frac{694449}{16384} \lambda_2^2
- \frac{64071}{8192} \lambda_3^2
- \frac{9477}{32} \zeta(3) C_F \lambda_1 \lambda_2^2
- \frac{9477}{128} \zeta(3) C_F \lambda_1 \lambda_3^2
\right. \right. \nonumber \\
&& \left. \left. ~~~~~~~~~~~~~~~
- \frac{37179}{256} \zeta(3) \lambda_1 \lambda_2
+ \frac{37179}{512} \zeta(3) \lambda_1 \lambda_3
+ \frac{12393}{256} \zeta(3) \lambda_2^2
+ \frac{37179}{4096} \zeta(3) \lambda_3^2
\right)
\right. \nonumber \\
&& \left. ~~~~~
+~ C_A^4 \left(
-~ \frac{192213}{1024} \lambda_1 \lambda_2^2
- \frac{192213}{4096} \lambda_1 \lambda_3^2
+ \frac{111537}{512} \zeta(3) \lambda_1 \lambda_2^2
+ \frac{111537}{2048} \zeta(3) \lambda_1 \lambda_3^2
\right)
\right. \nonumber \\
&& \left. ~~~~~
+ \left(
-~ \frac{247}{4096} C_F
+ \frac{1185}{1024} C_F^2 \lambda_2
- \frac{615}{1024} C_F^2 \lambda_3
+ \frac{1}{128} \zeta(2) \Nf T_F
+ \frac{3}{64} \zeta(3) C_F
\right)
\right. \nonumber \\
&& \left. ~~~~~
+~ \left[ C_A \left(
+~ \frac{75}{4096}
- \frac{315}{1024} C_F \lambda_2
\right)
+ C_A^2 \left(
+ \frac{315}{4096} \lambda_2
\right)
+ \frac{9}{1024} C_F
\right] \ln \left( \frac{g \sigma^\prime}{{\mu}^2} \right)
\right. \nonumber \\
&& \left. ~~~~~
-~ \frac{9}{4096} C_A \left( \ln \left( \frac{g \sigma^\prime}{{\mu}^2} 
\right) \right)^2  
\right] \frac{g^4 N_A \sigma^{\prime \, 2}}{\pi^4} ~+~ O(g^6)  
\label{effpotsig}
\end{eqnarray}
where  
\begin{equation}
\lambda_1 ~=~ [13 C_A-8 T_F \Nf]^{-1} ~~,~~
\lambda_2 ~=~ [35 C_A-16 T_F \Nf]^{-1} ~~,~~
\lambda_3 ~=~ [19 C_A-8 T_F \Nf]^{-1} 
\end{equation}
and $\mu$ is the renormalization scale which incorporates the usual factor of
$4\pi e^{-\gamma}$ into the $\mbox{MS}$ renormalization scale where $\gamma$ is
the Euler-Mascheroni constant. Minimizing $V(\sigma)$ with respect to the 
quantity $\sigma$ one discovers that the classical perturbative vacuum at 
$\langle \sigma \rangle$~$=$~$0$ is unstable and that there is a stable vacuum 
for a value of $\langle \sigma \rangle$~$\neq$~$0$. Estimates for the value of 
$\langle \sigma \rangle$ were given in \cite{8} by assuming that 
\begin{equation} 
\frac{dV(\sigma)}{d\sigma} ~=~ 0
\label{vsigcond}
\end{equation}
and then choosing the renormalization scale $\mu$ to be such that there were no
logarithms in the equation relating the coupling constant to the value of
$\langle \sigma \rangle$. Using the renormalization group properties relating
the coupling constant to the scale $\mu$ and hence the fundamental scale 
$\Lambda_{\mbox{\footnotesize{QCD}}}$ in the $\MSbar$ scheme, the estimate for
$\langle \sigma \rangle$ was obtained which was relatively stable to two loop
corrections. In \cite{8} a subsequent estimate was deduced for an effective
gluon mass. This will be illustrated in more detail in a later section.  

However, as indicated earlier this was essentially the classical or bare gluon
mass in the region of the stable minimum of the effective potential for
$\sigma$. A more appropriate quantity to examine would be a gluon mass where
some account of the quantum corrections were included. In this article we will
use the pole mass of the gluon as constructed from the LCO Lagrangian at two
loops building on the previous one loop analysis of \cite{25}. 

\sect{Two loop gluon pole mass.} 
In this section we construct the relation between the running gluon mass of the
LCO Lagrangian and the pole mass which is defined to be the pole of the one
particle irreducible gluon polarization tensor. In \cite{27} the simpler two 
dimensional Gross-Neveu model was studied and the relation between the 
analogous quantities was determined. The final mass estimates compared
favourably with the known exact mass gap. Whilst we will use \cite{27} as a
basis for the QCD computation there are significant differences aside from the
space-time dimensionality. The first is that $L^\sigma$ has more interactions
and basic fields as well as the gauge property. Second, and partly as a 
consequence of the previous point, it is not possible to fully construct the
gluon $2$-point function for all momenta and then deduce the pole of the
propagator. This is also due to the fact that not all relevant basic $2$-point 
two loop Feynman diagrams can be written in terms of closed known analytic
functions for all values of the momenta. To circumvent these difficulties we 
have followed the strategy and algorithm of a similar model in the context of 
the weak sector of the full standard model. Our approach is based on the series
of articles \cite{28,29,30,31} which applies the {\sc On-Shell} algorithm to 
the relation of the vector gauge boson poles masses in $\MSbar$ to their bare 
values. This package, \cite{28,29}, is designed to determine the value of two 
loop Feynman diagrams with massless propagators in addition to a propagator 
with a mass which is the on-shell value whose pole mass one is interested in. 
It uses dimensional regularization in $d$~$=$~$4$~$-$~$2\epsilon$ dimensions. 
One can extend the approach of \cite{28,29} to integrals with more than one 
scale by expanding in an appropriate ratio of masses which is assumed to be 
small. In our case this complication does not occur.

The {\sc On-Shell} package, \cite{28,29}, is written in the symbolic 
manipulation language {\sc Form}, \cite{32}, and for $L^\sigma$ we have 
generated the relevant one and two loop one particle irreducible Feynman 
diagrams using the {\sc Qgraf} package, \cite{33}. This is converted into a 
{\sc Form} readable format before applying the {\sc On-Shell} procedure to 
determine the value of each individual diagram when the external momentum is 
set to its on-shell value. For the LCO Lagrangian we are interested in there 
are $5$ one loop diagrams and $39$ two loop diagrams to evaluate.

The remaining issue is to construct the pole mass itself from the integral
contributing to the gluon $2$-point polarization. If we define the transverse
part of the correction to the polarization tensor by 
\begin{equation}
\Pi_{\mu\nu}(p) ~=~ \Pi(p^2,m^2) \left[ \eta_{\mu\nu} ~-~ 
\frac{p_\mu p_\nu}{p^2} \right]
\end{equation} 
where $p$ is the external momentum then the pole mass is defined to be that 
value of $p^2$ which is the solution to, \cite{30}, 
\begin{equation}
p^2 ~-~ m^2 ~-~ \Pi(p^2,m^2) ~=~ 0 ~. 
\end{equation} 
If we write the perturbative expansion of the transverse part of the 
polarization tensor as 
\begin{equation}
\Pi(p^2,m^2) ~=~ \sum_{n=1}^\infty \Pi_n(p^2,m^2) g^{2n}
\end{equation}
then to two loops one can solve the pole mass condition iteratively to obtain 
the pole mass, $s_p$, as, \cite{30}, 
\begin{equation} 
s_p ~=~ m^2 ~+~ \Pi_1(m^2,m^2) g^2 ~+~ \left( \Pi_2(m^2,m^2) ~+~ 
\Pi_1(m^2,m^2) \Pi^\prime_1(m^2,m^2) \right) g^4 ~+~ O(g^6)
\label{poledef}
\end{equation}  
where
\begin{equation}
\Pi^\prime_1(m^2,m^2) ~=~ \left. \frac{\partial~}{\partial p^2} \Pi_1(p^2,m^2)
\right|_{p^2=m^2} 
\end{equation} 
and here $m^2$ $=$ $m^2(\mu)$ is the running mass. The actual values of 
$\Pi_i(m^2,m^2)$ are obtained from the {\sc On-Shell} package, \cite{28,29}. In
determining the two loop part of (\ref{poledef}) from the one loop diagrams,
we have expanded the bare coupling constant and bare mass in terms of the 
renormalized variables before applying the one and two loop {\sc On-Shell}
routines. As a check that the final expression we obtain for the pole mass is
correct we note that first the full $2$-point function $\Pi(m^2,m^2)$ itself 
has to be finite at two loops after renormalization with the usual Landau gauge
renormalization constants, \cite{34,35,36,37}, and, \cite{8,9,10,22}, 
\begin{eqnarray}
Z_\xi^{-1} &=& 1 ~+~ \left( \frac{13}{6} C_A - \frac{4}{3} T_F \Nf \right) 
\frac{g^2}{16 \pi^2 \epsilon} \nonumber \\ 
&& +~ \left[ \left( 1464 C_A^2 T_F \Nf - 1365 C_A^3 - 384 C_A T_F^2 \Nf^2 
\right) \frac{1}{\epsilon^2} \right. \nonumber \\
&& \left. ~~~~~+~ \left( 5915 C_A^3 - 6032 C_A^2 T_F \Nf - 1248 C_A C_F T_F \Nf 
+ 1472 C_A T_F^2 \Nf^2 \right. \right. \nonumber \\
&& \left. \left. ~~~~~~~~~~~+~ 768 C_F T_F^2 \Nf^2 \right) \frac{1}{\epsilon} 
\right] \frac{g^4}{6144 \pi^4 (35 C_A-16 T_F \Nf)} ~+~ O(g^6) 
\end{eqnarray}
in the $\MSbar$ scheme. This is useful since it checks that the expansion of 
the one loop diagrams has been performed correctly when the coupling constant
and mass are replaced by their renormalized variables. The quantity 
$\Pi^\prime_1(m^2,m^2)$ is itself clearly finite by simple power counting. 
Finally, we arrive at our expression for the two loop $\MSbar$ pole mass using 
the LCO Lagrangian which, with $s_p$~$=$~$m^2_{\mbox{\footnotesize{LCO}}}$, is  
\begin{eqnarray}
m^2_{\mbox{\footnotesize{LCO}}} &=& \left[ 1 ~+~ \left( \left( 
\frac{287}{576} ~-~ \frac{3}{64} \ln \left( \frac{m^2(\mu)}{\mu^2} 
\right) ~-~ \frac{11\pi\sqrt{3}}{128} \right) C_A - \frac{1}{9} T_F \Nf \right) 
\frac{g^2}{\pi^2} \right. \nonumber \\
&& \left. ~~~~+~ \left( \left( \frac{1}{8} \zeta(3) - \frac{39}{256} 
+ \frac{1}{128} \ln \left( \frac{m^2(\mu)}{\mu^2} \right) \right) T_F \Nf C_F 
\right. \right. \nonumber \\
&& \left. \left. ~~~~~~~~~~~~+~ \left( - \frac{2801}{55296}
+ \frac{99}{1024} S_2 - \frac{647}{6912} \ln \left( \frac{m^2(\mu)}{\mu^2} 
\right) + \frac{3}{512} \left[ \ln \left( \frac{m^2(\mu)}{\mu^2} \right) 
\right]^2 \right. \right. \right. \nonumber \\
&& \left. \left. \left. ~~~~~~~~~~~~~~~~~~~~+~ \frac{379}{4608} \zeta(2)
- \frac{1}{8} \zeta(3) \right) T_F \Nf C_A \right. \right. \nonumber \\
&& \left. \left. ~~~~~~~~~~~~+~ \left( \frac{3}{32} S_2 - \frac{11}{3456} 
+ \frac{11}{1536} \ln \left( \frac{m^2(\mu)}{\mu^2} \right) \right) \sqrt{3} 
T_F \Nf C_A \right. \right. \nonumber \\
&& \left. \left. ~~~~~~~~~~~~+~ \left( - \frac{7}{432}
+ \frac{1}{54} \ln \left( \frac{m^2(\mu)}{\mu^2} \right)
- \frac{1}{72} \zeta(2) + \frac{\pi^2}{144} \right) T_F^2 \Nf^2 \right. \right.
\nonumber \\  
&& \left. \left. ~~~~~~~~~~~~+~ \left( \frac{3}{2048} - \frac{9}{2048 } \ln 
\left( \frac{m^2(\mu)}{\mu^2} \right) \right) C_F C_A \right. \right. 
\nonumber \\
&& \left. \left. ~~~~~~~~~~~~+~ \left( - \frac{105}{2048} + \frac{315}{2048} 
\ln \left( \frac{m^2(\mu)}{\mu^2} \right) \right) 
\frac{C_F C_A^2}{[35 C_A - 16 T_F \Nf]} \right. \right. \nonumber \\
&& \left. \left. ~~~~~~~~~~~~+~ \left( \frac{9737}{24576}
- \frac{3069}{8192} S_2 + \frac{11461}{221184} \ln \left( 
\frac{m^2(\mu)}{\mu^2} \right) \right. \right. \right. \nonumber \\
&& \left. \left. \left. ~~~~~~~~~~~~~~~~~~~~-~ \frac{51}{8192} \left[ \ln 
\left( \frac{m^2(\mu)}{\mu^2} \right) \right]^2 - \frac{59}{2304} \zeta(2)
+ \frac{231}{8192} \zeta(3) \right) C_A^2 \right. \right. \nonumber \\
&& \left. \left. ~~~~~~~~~~~~+~ \left( \frac{105}{8192} - \frac{315}{8192} \ln 
\left( \frac{m^2(\mu)}{\mu^2} \right) \right) 
\frac{C_A^3}{[35 C_A - 16 T_F \Nf]} \right. \right. \nonumber \\
&& \left. \left. ~~~~~~~~~~~~+~ \left( - \frac{1413}{32768} S_2 
- \frac{12503}{221184} + \frac{77}{24576} \ln \left( \frac{m^2(\mu)}{\mu^2} 
\right) \right) \sqrt{3} \pi C_A^2 \right. \right. \nonumber \\
&& \left. \left. ~~~~~~~~~~~~+~ \frac{17\pi^2}{2304} C_A^2  \right) 
\frac{g^4}{\pi^4} ~+~ O(g^6) \right] m^2(\mu)  
\label{lcomass} 
\end{eqnarray}
where $S_2$ $=$ $(4 \sqrt{3}/3) Cl_2(\pi/3)$ and $Cl_2(x)$ is the Clausen 
function. 

Another check on the symbolic manipulation routines we have written was to 
consider the Curci-Ferrari Lagrangian, \cite{19}, in the Landau gauge which is 
effectively QCD with a gluon mass included by hand. The corresponding 
Lagrangian is  
\begin{equation} 
L^{\mbox{\footnotesize{mQCD}}} ~=~ L^{\mbox{\footnotesize{QCD}}} ~+~ \half 
m^2 A^a_\mu A^{a\,\mu} ~-~ \alpha m^2 \bar{c}^a c^a 
\label{cflag} 
\end{equation} 
where we have included the ghost mass term for completeness. As the one loop
pole mass for this theory was given in \cite{26,25}, it does not require much
more effort to produce the two loop $\MSbar$ correction using the same symbolic
manipulation programmes. In this case the same {\sc Qgraf} output is used but 
with a null $\sigma$ vertex and the usual quartic gluon interaction. The same
check that the $2$-point function is finite was satisfied. By contrast to
(\ref{lcomass}) we find that the pole mass for (\ref{cflag}) is, with 
$s_p$~$=$~$m^2_{\mbox{\footnotesize{CF}}}$, 
\begin{eqnarray}
m^2_{\mbox{\footnotesize{CF}}} &=& \left[ 1 ~+~ \left( \left( 
\frac{313}{576} ~-~ \frac{35}{192} \ln \left( \frac{m^2(\mu)}{\mu^2} 
\right) ~-~ \frac{11\pi\sqrt{3}}{128} \right) C_A \right. \right. \nonumber \\
&& \left. \left. ~~~~~~~~~~~~+~ \left( \frac{1}{12} \ln \left( 
\frac{m^2(\mu)}{\mu^2} \right) - \frac{5}{36} \right) T_F \Nf \right) 
\frac{g^2}{\pi^2} \right. \nonumber \\  
&& \left. ~~~~+~ \left( \left( \frac{1}{8} \zeta(3) - \frac{119}{768} 
+ \frac{1}{64} \ln \left( \frac{m^2(\mu)}{\mu^2} \right) \right) T_F \Nf C_F 
\right. \right. \nonumber \\
&& \left. \left. ~~~~~~~~~~~~+~ \left( - \frac{20335}{165888}
+ \frac{297}{1024} S_2 + \frac{13}{13824} \ln \left( \frac{m^2(\mu)}{\mu^2} 
\right) + \frac{95}{4608} \left[ \ln \left( \frac{m^2(\mu)}{\mu^2} \right) 
\right]^2 \right. \right. \right. \nonumber \\
&& \left. \left. \left. ~~~~~~~~~~~~~~~~~~~~+~ \frac{91}{1536} \zeta(2)
- \frac{1}{8} \zeta(3) \right) T_F \Nf C_A \right. \right. \nonumber \\
&& \left. \left. ~~~~~~~~~~~~+~ \left( \frac{3}{32} S_2 + \frac{11}{13824} 
- \frac{11}{2304} \ln \left( \frac{m^2(\mu)}{\mu^2} \right) \right) \sqrt{3} 
T_F \Nf C_A \right. \right. \nonumber \\
&& \left. \left. ~~~~~~~~~~~~+~ \left( - \frac{5}{648}
+ \frac{7}{432} \ln \left( \frac{m^2(\mu)}{\mu^2} \right)
- \frac{1}{144} \left[ \ln \left( \frac{m^2(\mu)}{\mu^2} \right) \right]^2 
+ \frac{\pi^2}{144} \right) T_F^2 \Nf^2 \right. \right. \nonumber \\  
&& \left. \left. ~~~~~~~~~~~~+~ \left( \frac{163265}{331776}
- \frac{5643}{8192} S_2 - \frac{11057}{110592} \ln \left( 
\frac{m^2(\mu)}{\mu^2} \right) \right. \right. \right. \nonumber \\
&& \left. \left. \left. ~~~~~~~~~~~~~~~~~~~~-~ \frac{875}{73728} \left[ \ln 
\left( \frac{m^2(\mu)}{\mu^2} \right) \right]^2 - \frac{51}{2048} \zeta(2)
+ \frac{231}{8192} \zeta(3) \right) C_A^2 \right. \right. \nonumber \\
&& \left. \left. ~~~~~~~~~~~~+~ \left( - \frac{1413}{32768} S_2 
- \frac{13933}{221184} + \frac{1661}{73728} \ln \left( \frac{m^2(\mu)}{\mu^2} 
\right) \right) \sqrt{3} \pi C_A^2 \right. \right. \nonumber \\
&& \left. \left. ~~~~~~~~~~~~+~ \frac{17\pi^2}{2304} C_A^2 \right) 
\frac{g^4}{\pi^4} ~+~ O(g^6) \right] m^2(\mu) ~. 
\label{cfplm} 
\end{eqnarray}
Although the Lagrangian for QCD in the non-linear Curci-Ferrari gauge formally 
differs from that for the Landau gauge in relation to the ghost gluon 
interaction term for $\alpha$~$=$~$0$, we have checked that the same pole mass 
emerges as (\ref{cfplm}) for the usual covariant Landau gauge fixing.  

\sect{Analysis.} 
The next stage of our analysis is to produce an estimate for the pole mass. In 
\cite{27} another method of estimating the Gross-Neveu mass gap was used. This 
required knowledge of the full $2$-point function as a function of the external
momentum. This approach is not available to us for QCD since the technology 
does not exist to compute the gluon $2$-point function {\em exactly} as a 
function of momentum. Instead we simply extend the argument of \cite{25} to two
loops. In \cite{8} an estimate for an effective gluon mass was obtained by 
examining the location of the minimum of $V(\sigma)$ using (\ref{vsigcond}). 
The effective gluon mass is essentially the bare mass of $L^\sigma$. It does 
not take account of quantum corrections. In \cite{25} it was argued that a more
appropriate quantity to estimate from the effective potential was 
$m^2_{\mbox{\footnotesize{LCO}}}$ itself. Specifically the quantity 
$V^{\mbox{\footnotesize{eff}}}(m^2_{\mbox{\footnotesize{LCO}}})$ was 
constructed by inverting the one loop part of (\ref{lcomass}) to obtain 
$m^2(\mu)$ as a function of $m^2_{\mbox{\footnotesize{LCO}}}$ before 
substituting for $m^2(\mu)$ using the relation with $\langle \sigma \rangle$ 
\begin{equation}
m^2(\mu) ~=~ \frac{9 N_{\! A} \langle \sigma \rangle}{[13 C_A - 8 T_F \Nf] 
g\xi(g)} ~. 
\end{equation} 
Thus we find the potential, truncated to one loop, is 
\begin{eqnarray}
V^{\mbox{\footnotesize{eff}}} \left(m^2_{\mbox{\footnotesize{LCO}}}\right) &=& 
\left[ \frac{9}{2} \lambda_1 + \left( - \frac{29}{128} + \frac{3}{64} 
\ln \left( \frac{m^2_{\mbox{\footnotesize{LCO}}}}{\mu^2} \right) 
- \frac{207}{32}C_F \lambda_2 + \frac{117}{32}C_F \lambda_3 \right. \right. 
\nonumber \\
&& \left. \left. ~~~~~~~~~~~+~ C_A \left( - \frac{351}{8}C_F\lambda_1 \lambda_2
+\frac{351}{16}C_F \lambda_1 \lambda_3 - \frac{183}{64} \lambda_1
\right. \right. \right. \nonumber \\
&& \left. \left. \left. ~~~~~~~~~~~~~~~~~~~~~~-~ \frac{249}{128} \lambda_2 
+ \frac{27}{64} \lambda_3 + \frac{99}{128}\pi \sqrt{3} \lambda_1 \right) 
\right. \right. \nonumber \\ 
&& \left. \left. ~~~~~~~~~~~+~ C_A^2 \left( - \frac{81}{16} \lambda_1 \lambda_2
+ \frac{81}{32} \lambda_1 \lambda_3 \right) ~+~ \frac{27}{64}C_A \lambda_1 \ln 
\left( \frac{m^2_{\mbox{\footnotesize{LCO}}}}{\mu^2} \right) \right) 
\frac{g^2}{\pi^2} \right. \nonumber \\
&& \left. ~+~ O(g^4) \frac{}{} \right] 
\frac{\left( 13 C_A - 8 T_F \Nf \right)^2}{81 N_{\! A}} g^2 \xi^2(g) 
m^4_{\mbox{\footnotesize{LCO}}} ~.  
\end{eqnarray} 
Using the minimization criterion, \cite{25},  
\begin{equation}
\frac{d V^{\mbox{\footnotesize{eff}}}(m^2_{\mbox{\footnotesize{LCO}}})}
{d m^2_{\mbox{\footnotesize{LCO}}}} ~=~ 0 
\label{masscond}
\end{equation}  
the following condition  
\begin{eqnarray}
0 &=& \left[ \frac{9}{2} \lambda_1 + \left( - \frac{13}{64} + \frac{3}{64} 
\ln \left( \frac{m^2_{\mbox{\footnotesize{LCO}}}}{\mu^2} \right) 
- \frac{207}{32}C_F \lambda_2 + \frac{117}{32}C_F \lambda_3 \right. \right. 
\nonumber \\
&& \left. \left. ~~~~~~~~~~~+~ C_A \left( - \frac{351}{8}C_F\lambda_1 \lambda_2
+\frac{351}{16}C_F \lambda_1 \lambda_3 - \frac{339}{128} \lambda_1
\right. \right. \right. \nonumber \\
&& \left. \left. \left. ~~~~~~~~~~~~~~~~~~~~~~-~ \frac{249}{128} \lambda_2 
+ \frac{27}{64} \lambda_3 + \frac{99}{128}\pi \sqrt{3} \lambda_1 \right) 
\right. \right. \nonumber \\ 
&& \left. \left. ~~~~~~~~~~~+~ C_A^2 \left( - \frac{81}{16} \lambda_1 \lambda_2
+ \frac{81}{32} \lambda_1 \lambda_3 \right) ~+~ \frac{27}{64}C_A \lambda_1 \ln 
\left( \frac{m^2_{\mbox{\footnotesize{LCO}}}}{\mu^2} \right) \right) 
\frac{g^2}{\pi^2} \right. \nonumber \\
&& \left. ~+~ O(g^4) \frac{}{} \right] 
\frac{\left( 13 C_A - 8 T_F \Nf \right)^2}{81 N_{\! A}} g^2 \xi^2(g) 
m^2_{\mbox{\footnotesize{LCO}}} 
\label{gap1}
\end{eqnarray} 
emerged where we have not included the explicit expansion of $\xi(g)$. This is
because it would introduce an unnecessary truncation error into the estimates
for the pole mass. Ignoring the trivial solution of 
$m^2_{\mbox{\footnotesize{LCO}}}$~$=$~$0$ which corresponds to the unstable 
vacuum, the non-trivial condition determines the pole mass estimate at one 
loop. To solve this the renormalization scale $\mu$ was parametrically related 
to $m^2_{\mbox{\footnotesize{LCO}}}$ by 
$m^2_{\mbox{\footnotesize{LCO}}}$~$=$~$s\mu^2$ which leaves a parametric 
relation between the running coupling constant and 
$m^2_{\mbox{\footnotesize{LCO}}}$ 
\begin{eqnarray} 
y &=& 36 C_A \left( 16 T_F \Nf - 35 C_A \right) \left[ \left( 3465 \pi \sqrt{3}
+ 4620 \ln(s) - 25690 \right) C_A^2 - 864 C_F T_F \Nf \right. \nonumber \\
&& \left. ~~~~~~~~~~~~~~~~~~~~~~~~~~~~~~~~+~ \left( 19240 - 1584 \pi \sqrt{3} 
- 3792 \ln(s) \right) C_A T_F \Nf \right. \nonumber \\
&& \left. ~~~~~~~~~~~~~~~~~~~~~~~~~~~~~~~~+~ \left( 768 \ln(s) - 3328 \right) 
T_F^2 \Nf^2 \right]^{-1}
\end{eqnarray}  
where $y$ $=$ $C_A g^2/(16\pi^2)$. However, from the one loop $\beta$-function
the coupling constant can be related to the fundamental scale 
$\Lambda_{\mbox{\footnotesize{{$\MSbar$}}}}$ using  
\begin{equation} 
\frac{g^2(\mu)}{16\pi^2} ~=~ \left[ \beta_0 \ln \left[ \frac{\mu^2}
{\Lambda^2_{\mbox{\footnotesize{$\MSbar$}}}}\right] \right]^{-1}  
\label{beta1}
\end{equation} 
where 
\begin{equation} 
\beta_0 ~=~ \frac{11}{3} C_A ~-~ \frac{4}{3} T_F \Nf ~. 
\end{equation} 
Hence, the $\mu$ independent estimate for the pole mass emerged, \cite{25},  
\begin{eqnarray}
m_{\mbox{\footnotesize{LCO}}} &=& 
\Lambda^{(\Nf)}_{\mbox{\footnotesize{$\MSbar$}}} \exp \left[ \,-\, \left( 
\left( 3465 \pi \sqrt{3} - 25690 \right) C_A^2 - 864 C_F T_F \Nf \right. 
\right. \nonumber \\
&& \left. \left. ~~~~~~~~~~~~~~~~~~~+~ \left( 19240 - 1584 \pi \sqrt{3} \right) 
C_A T_F \Nf ~-~ 3328 T_F^2 \Nf^2 \right) \right. \nonumber \\
&& \left. ~~~~~~~~~~~~~~ \left( \frac{}{} \! 24 \left( 11 C_A - 4 T_F \Nf 
\right) \left( 35 C_A - 16 T_F \Nf \right) \right)^{-1} \right] ~. 
\end{eqnarray}

Equipped with the two loop pole mass of (\ref{lcomass}), we now repeat the 
above analysis by first inverting $m^2_{\mbox{\footnotesize{LCO}}}$ to obtain 
$m^2(\mu)$ at two loops as a function of $m^2_{\mbox{\footnotesize{LCO}}}$.
Substituting this into (\ref{effpotsig}) and truncating at two loops, we find
\begin{eqnarray}
V^{\mbox{\footnotesize{eff}}} \left(m^2_{\mbox{\footnotesize{LCO}}}\right) &=& 
\left[ \frac{9}{2} \lambda_1 + \left( - \frac{29}{128} - \frac{207}{32}C_F 
\lambda_2 + \frac{117}{32}C_F \lambda_3 \right. \right. \nonumber \\
&& \left. \left. ~~~~~~~~~~+~ C_A \left( - \frac{351}{8}C_F\lambda_1 \lambda_2
+\frac{351}{16}C_F \lambda_1 \lambda_3 - \frac{183}{64} \lambda_1
\right. \right. \right. \nonumber \\
&& \left. \left. \left. ~~~~~~~~~~~~~~~~~~~~~-~ \frac{249}{128} \lambda_2 
+ \frac{27}{64} \lambda_3 + \frac{99}{128}\pi \sqrt{3} \lambda_1 \right) 
\right. \right. \nonumber \\ 
&& \left. \left. ~~~~~~~~~~+~ C_A^2 \left( - \frac{81}{16} \lambda_1 \lambda_2
+ \frac{81}{32} \lambda_1 \lambda_3 \right) + \frac{3}{64} \ln \left( 
\frac{m^2_{\mbox{\footnotesize{LCO}}}}{\mu^2} \right) \right. \right. 
\nonumber \\
&& \left. \left. ~~~~~~~~~~+~ \frac{27}{64}C_A \lambda_1 \ln \left( 
\frac{m^2_{\mbox{\footnotesize{LCO}}}}{\mu^2} \right) 
\right) \frac{g^2}{\pi^2} \right. \nonumber \\ 
&& \left. +~ \left( \Nf T_F \left( - \frac{71}{1152} - \frac{1}{128} \zeta(2)
+ \frac{1}{128} \pi^2 \right) \right. \right. \nonumber \\
&& \left. \left. ~~~~~~
       +~ C_F C_A   \left(
          + \frac{567}{256} \lambda_1
          - \frac{23067}{2048} \lambda_2
          + \frac{27133}{4096} \lambda_3
          - \frac{117}{64} \zeta(3) \lambda_1
\right. \right. \right. \nonumber \\
&& \left. \left. \left. ~~~~~~~~~~~~~~~~~~~~ 
          + \frac{585}{16} \zeta(3) \lambda_2
          - \frac{4881}{256} \zeta(3) \lambda_3
          \right)
\right. \right. \nonumber \\
&& \left. \left. ~~~~~~ 
       +~ C_F C_A \pi   \left(
          - \frac{2277}{2048} \sqrt{3} \lambda_2
          + \frac{1287}{2048} \sqrt{3} \lambda_3
          \right) \right. \right. \nonumber \\
&& \left. \left. ~~~~~~
       +~ C_F C_A^2   \left(
          - \frac{127287}{2048} \lambda_1 \lambda_2
          + \frac{16029}{512} \lambda_1 \lambda_3
          + \frac{72801}{2048} \lambda_2^2 
          + \frac{11583}{2048} \lambda_3^2
\right. \right. \right. \nonumber \\
&& \left. \left. \left. ~~~~~~~~~~~~~~~~~~~~ 
          + \frac{3159}{16} \zeta(3) \lambda_1 \lambda_2
          - \frac{3159}{32} \zeta(3) \lambda_1 \lambda_3
\right. \right. \right. \nonumber \\
&& \left. \left. \left. ~~~~~~~~~~~~~~~~~~~~ 
          - \frac{1053}{16} \zeta(3) \lambda_2^2
          - \frac{3159}{256} \zeta(3) \lambda_3^2
          \right) \right. \right. \nonumber \\
&& \left. \left. ~~~~~~
       +~ C_F C_A^2 \pi   \left(
          - \frac{3861}{512} \sqrt{3} \lambda_1 \lambda_2
          + \frac{3861}{1024} \sqrt{3} \lambda_1 \lambda_3
          \right) \right. \right. \nonumber \\
&& \left. \left. ~~~~~~
       +~ C_F C_A^3   \left(
          + \frac{34749}{256} \lambda_1 \lambda_2^2 
          + \frac{34749}{1024} \lambda_1 \lambda_3^2 
\right. \right. \right. \nonumber \\
&& \left. \left. \left. ~~~~~~~~~~~~~~~~~~~~ 
          - \frac{9477}{32} \zeta(3) \lambda_1 \lambda_2^2 
          - \frac{9477}{128} \zeta(3) \lambda_1 \lambda_3^2
          \right) \right. \right. \nonumber \\
&& \left. \left. ~~~~~~
       +~ C_F   \left(
          - \frac{1015}{4096}
          + \frac{3}{16} \zeta(3)
          \right) \right. \right. \nonumber \\
&& \left. \left. ~~~~~~
       +~ C_F^2 C_A   \left(
          - \frac{1053}{64} \lambda_1 \lambda_2
          + \frac{1053}{128} \lambda_1 \lambda_3
          - \frac{5409}{1024} \lambda_2^2
          + \frac{1053}{1024} \lambda_3^2
          \right) \right. \right. \nonumber \\
&& \left. \left. ~~~~~~
       +~ C_F^2 C_A^2   \left(
          + \frac{3159}{128} \lambda_1 \lambda_2^2
          + \frac{3159}{512} \lambda_1 \lambda_3^2
          \right) \right. \right. \nonumber \\
&& \left. \left. ~~~~~~
       +~ C_F^2   \left(
          + \frac{1185}{1024} \lambda_2
          - \frac{615}{1024} \lambda_3
          \right) \right. \right. \nonumber \\
&& \left. \left. ~~~~~~
       +~ C_A   \left(
          + \frac{9709}{73728}
          + \frac{891}{4096} S_2 
          + \frac{137}{2048} \zeta(2)
          - \frac{3}{16} \zeta(3)
          \right) \right. \right. \nonumber \\
&& \left. \left. ~~~~~~
       +~ C_A \pi   \left(
          - \frac{605}{12288} \sqrt{3}
          + \frac{27}{256} \sqrt{3} S_2 \right)  
       ~+~ C_A \pi^2   \left(
          + \frac{13}{1024}
          \right) \right. \right. \nonumber \\
&& \left. \left. ~~~~~~
       +~ C_A^2   \left(
          + \frac{8019}{4096} S_2 \lambda_1
          - \frac{19845}{16384} \lambda_1
          + \frac{389957}{16384} \lambda_2
          - \frac{193687}{16384} \lambda_3
\right. \right. \right. \nonumber \\
&& \left. \left. \left. ~~~~~~~~~~~~~~~~~ 
          - \frac{2631}{4096} \zeta(2) \lambda_1
          + \frac{12897}{8192} \zeta(3) \lambda_1
\right. \right. \right. \nonumber \\
&& \left. \left. \left. ~~~~~~~~~~~~~~~~~ 
          - \frac{6885}{256} \zeta(3) \lambda_2
          + \frac{116115}{8192} \zeta(3) \lambda_3
          \right) \right. \right. \nonumber \\
&& \left. \left. ~~~~~~
       +~ C_A^2 \pi   \left(
          - \frac{32211}{32768} \sqrt{3} S_2 \lambda_1
          - \frac{1193}{8192} \sqrt{3} \lambda_1
          - \frac{2739}{8192} \sqrt{3} \lambda_2
          + \frac{297}{4096} \sqrt{3} \lambda_3
          \right) \right. \right. \nonumber \\
&& \left. \left. ~~~~~~
       +~ C_A^2 \pi^2   \left(
          + \frac{2217}{32768} \lambda_1
          \right) \right. \right. \nonumber \\
&& \left. \left. ~~~~~~
       +~ C_A^3   \left(
          + \frac{1050543}{8192} \lambda_1 \lambda_2
          - \frac{32859}{512} \lambda_1 \lambda_3
          - \frac{694449}{16384} \lambda_2^2
          - \frac{64071}{8192} \lambda_3^2
\right. \right. \right. \nonumber \\
&& \left. \left. \left. ~~~~~~~~~~~~~~~~~ 
          - \frac{37179}{256} \zeta(3) \lambda_1 \lambda_2
          + \frac{37179}{512} \zeta(3) \lambda_1 \lambda_3
\right. \right. \right. \nonumber \\
&& \left. \left. \left. ~~~~~~~~~~~~~~~~~ 
          + \frac{12393}{256} \zeta(3) \lambda_2^2
          + \frac{37179}{4096} \zeta(3) \lambda_3^2
          \right) \right. \right. \nonumber \\
&& \left. \left. ~~~~~~
       +~ C_A^3 \pi   \left(
          - \frac{891}{1024} \sqrt{3} \lambda_1 \lambda_2
          + \frac{891}{2048} \sqrt{3} \lambda_1 \lambda_3
          \right) \right. \right. \nonumber \\
&& \left. \left. ~~~~~~
       +~ C_A^4   \left(
          - \frac{192213}{1024} \lambda_1 \lambda_2^2
          - \frac{192213}{4096} \lambda_1 \lambda_3^2
\right. \right. \right. \nonumber \\
&& \left. \left. \left. ~~~~~~~~~~~~~~~~~ 
          + \frac{111537}{512} \zeta(3) \lambda_1 \lambda_2^2
          + \frac{111537}{2048} \zeta(3) \lambda_1 \lambda_3^2
          \right) \right. \right. \nonumber \\
&& \left. \left. ~~~~~~
+~ \ln \left( \frac{m_{\mbox{\footnotesize{LCO}}}^2}{\mu^2} \right) \Nf T_F
\left(
          + \frac{1}{32}
          \right) \right. \right. \nonumber \\
&& \left. \left. ~~~~~~
+~ \ln \left( \frac{m_{\mbox{\footnotesize{LCO}}}^2}{\mu^2} \right) C_F C_A
\left(
          - \frac{153}{2048} \lambda_1
          - \frac{117}{128} \lambda_2
          + \frac{351}{1024} \lambda_3
          \right) \right. \right. \nonumber \\
&& \left. \left. ~~~~~~
+~ \ln \left( \frac{m_{\mbox{\footnotesize{LCO}}}^2}{\mu^2} \right) C_F C_A^2
\left(
          - \frac{11259}{2048} \lambda_1 \lambda_2
          + \frac{1053}{512} \lambda_1 \lambda_3
          \right) \right. \right. \nonumber \\
&& \left. \left. ~~~~~~
+~ \ln \left( \frac{m_{\mbox{\footnotesize{LCO}}}^2}{\mu^2} \right) C_F
\left(
          + \frac{9}{512}
          \right) 
~+~ \ln \left( \frac{m_{\mbox{\footnotesize{LCO}}}^2}{\mu^2} \right) C_A
\left(
          - \frac{511}{4096}
          \right) \right. \right. \nonumber \\
&& \left. \left. ~~~~~~
+~ \ln \left( \frac{m_{\mbox{\footnotesize{LCO}}}^2}{\mu^2} \right) C_A \pi
\left(
          + \frac{33}{2048} \sqrt{3}
          \right) \right. \right. \nonumber \\
&& \left. \left. ~~~~~~
+~ \ln \left( \frac{m_{\mbox{\footnotesize{LCO}}}^2}{\mu^2} \right) C_A^2 
\left(
          + \frac{657}{8192} \lambda_1
          - \frac{27}{256} \lambda_2
          + \frac{81}{2048} \lambda_3
          \right) \right. \right. \nonumber \\
&& \left. \left. ~~~~~~
+~ \ln \left( \frac{m_{\mbox{\footnotesize{LCO}}}^2}{\mu^2} \right) C_A^2 \pi 
\left(
          - \frac{99}{4096} \sqrt{3} \lambda_1
          \right) \right. \right. \nonumber \\
&& \left. \left. ~~~~~~
+~ \ln \left( \frac{m_{\mbox{\footnotesize{LCO}}}^2}{\mu^2} \right) C_A^3 
\left(
          - \frac{1053}{8192} \lambda_1 \lambda_2
          + \frac{243}{1024} \lambda_1 \lambda_3
          \right) \right. \right. \nonumber \\
&& \left. \left. ~~~~~~
+~ \left( \ln \left( \frac{m_{\mbox{\footnotesize{LCO}}}^2}{\mu^2} \right) 
\right)^2 C_A
\left(
          + \frac{9}{1024}
          \right)
\right) \frac{g^4}{\pi^4} \right. \nonumber \\
&& \left. ~~~~~+~ O(g^6) \right] 
\frac{\left( 13 C_A - 8 T_F \Nf \right)^2}{81 N_{\! A}} g^2 \xi^2(g) 
m^4_{\mbox{\footnotesize{LCO}}} 
\end{eqnarray}
where again we have not substituted for $\xi(g)$. As this general expression is
rather cumbersome, in order to illustrate the two loop analysis we concentrate 
for the moment on the case of $SU(3)$ with $\Nf$~$=$~$0$ when we simply have 
\begin{eqnarray}
\left. V^{\mbox{\footnotesize{eff}}} 
\left(m^2_{\mbox{\footnotesize{LCO}}}\right) 
\right|_{\mbox{\footnotesize{$SU(3)$}}}^{\mbox{\footnotesize{$\Nf$$=$$0$}}} &=& 
\left[ \frac{3}{26} 
+ \left( 99 \sqrt{3} \pi 
   + 132 \ln \left( \frac{m_{\mbox{\footnotesize{LCO}}}^2}{\mu^2} \right) 
   - 800 \right) \frac{g^2}{1664 \pi^2} \right. \nonumber \\
&& \left. ~+~ \left(1038312 \sqrt{3} \pi \ln \left( 
\frac{m_{\mbox{\footnotesize{LCO}}}^2}{\mu^2} \right) 
   + 2174607 \sqrt{3} \pi S_2
   - 4831320 \sqrt{3} \pi 
\right. \right. \nonumber \\
&& \left. \left. ~~~~~~~+~ 640224 \left( \ln \left( 
\frac{m_{\mbox{\footnotesize{LCO}}}^2}{\mu^2} \right) \right)^2 
   - 8656704 \ln \left( \frac{m_{\mbox{\footnotesize{LCO}}}^2}{\mu^2} \right) 
\right. \right. \nonumber \\
&& \left. \left. ~~~~~~~-~ 1752192 \zeta(3) 
   + 1516143 \pi^2 
   + 26815536 S_2 
\right. \right. \nonumber \\
&& \left. \left. ~~~~~~~+~ 2936668 \frac{}{} \right) 
\frac{g^4}{24281088 \pi^4} ~+~ O(g^6) \right] \frac{169}{72} g^2 \xi^2(g) 
m^4_{\mbox{\footnotesize{LCO}}} ~. \nonumber \\ 
\end{eqnarray}
Solving (\ref{masscond}) as before and discarding the trivial solution yields
the two loop correction to (\ref{gap1}) for $SU(3)$ Yang-Mills theory,   
\begin{eqnarray}
0 &=& \frac{3}{13} ~+~ \left( 99 \sqrt{3} \pi 
   + 132 \ln \left( \frac{m_{\mbox{\footnotesize{LCO}}}^2}{\mu^2} \right) 
   - 734 \right)
\frac{g^2}{832 \pi^2} \nonumber \\
&& +~ \left( 1038312 \sqrt{3} \pi  
\ln \left( \frac{m_{\mbox{\footnotesize{LCO}}}^2}{\mu^2} \right) 
   + 2174607 \sqrt{3} \pi S_2
   - 4312164 \sqrt{3} \pi \right. \nonumber \\
&& \left. ~~~~~+~ 640224 \left( \ln \left( 
\frac{m_{\mbox{\footnotesize{LCO}}}^2}{\mu^2} \right) \right)^2  
   - 8016480 \ln \left( \frac{m_{\mbox{\footnotesize{LCO}}}^2}{\mu^2} \right) 
   - 1752192 \zeta(3) \right. \nonumber \\
&& \left. ~~~~~+~ 1516143 \pi^2 
   + 26815536 S_2
   - 1391684 \frac{}{} \right) \frac{g^4}{12140544 \pi^4} ~+~ O(g^6) ~. 
\label{gap2}
\end{eqnarray}
In analysing this along the lines of the one loop case, it transpires that the 
resulting two loop correction for the $m^2_{\mbox{\footnotesize{LCO}}}$ 
estimate is not $\mu$ independent. Therefore, we choose to return to the 
procedure of \cite{8} and select the scale $\mu^2$ so as to remove the 
logarithms in (\ref{gap2}). This fixes $g(\mu)$ to a particular numerical value
but using it the mass scale is recovered from the two loop extension of 
(\ref{beta1})  
\begin{equation} 
\frac{g^2(\mu)}{16\pi^2} ~=~ \left[ \beta_0 \ln \left[ \frac{\mu^2}
{\Lambda^2_{\mbox{\footnotesize{$\MSbar$}}}}\right] \right]^{-1} \left[ 1 ~-~ 
\beta_1 \left[\beta_0^2 \ln \left[ \frac{\mu^2}
{\Lambda^2_{\mbox{\footnotesize{$\MSbar$}}}}\right] \right]^{-1}  
\ln \left[ \ln \left[ \frac{\mu^2}
{\Lambda^2_{\mbox{\footnotesize{$\MSbar$}}}}\right] \right] \right] 
\end{equation} 
where
\begin{equation} 
\beta_1 ~=~ \frac{34}{3} C_A^2 ~-~ 4 C_F T_F \Nf ~-~ \frac{20}{3} 
C_A T_F \Nf ~.
\end{equation} 
{\begin{table}[ht] 
\begin{center} 
\begin{tabular}{|c||c|c|} 
\hline
$N_{\!f}$ & \mbox{$1$ loop} & \mbox{$2$ loop} \\ 
\hline
 0 & 2.10 & 2.13 \\ 
 2 & 1.74 & 2.21 \\ 
 3 & 1.55 & 2.32 \\ 
\hline
\end{tabular} 
\end{center} 
\begin{center} 
{Table 1. One and two loop estimates of $m_{\mbox{\footnotesize{LCO}}}
/\Lambda^{(\Nf)}_{\mbox{\footnotesize{$\MSbar$}}}$ for $SU(3)$.}
\end{center} 
\end{table}}  

We have obtained estimates for both groups $SU(2)$ and $SU(3)$ for several 
quark flavours. These are summarized in Tables $1$ and $2$. Several features 
emerge. First, for Yang-Mills interestingly the two loop correction is less 
than $2\%$ percent of the one loop value which suggests that our approximation 
is reliable. Unfortunately when quarks are included for both colour groups the
situation is different with the two loop estimates being significantly larger
than the one loop ones. Though for $SU(3)$ they are of a similar size as the 
Yang-Mills value. Given the stability of the two loop results for 
$\Nf$~$=$~$0$ compared with $\Nf$~$\neq$~$0$, it would suggest that the 
analysis when quarks are present is lacking some stabilising ingredient. One
possibility is that for full QCD one actually requires quark masses to be
included.  

{\begin{table}[ht] 
\begin{center} 
\begin{tabular}{|c||c|c|} 
\hline
$N_{\!f}$ & \mbox{$1$ loop} & \mbox{$2$ loop} \\ 
\hline
 0 & 2.10 & 2.13 \\ 
 2 & 1.54 & 2.29 \\ 
 3 & 1.24 & 2.58 \\ 
\hline
\end{tabular} 
\end{center} 
\begin{center} 
{Table 2. One and two loop estimates of $m_{\mbox{\footnotesize{LCO}}}
/\Lambda^{(\Nf)}_{\mbox{\footnotesize{$\MSbar$}}}$ for $SU(2)$.}
\end{center} 
\end{table}}  

\sect{Discussion.} 
We have produced estimates for the gluon pole mass in QCD from the local 
composite operator method which systematically introduces an extra scalar field
coupled to the gluon mass operator into the Lagrangian. The one loop 
renormalization scale independent estimate of \cite{25} is stable to the two 
loop corrections for Yang-Mills theory but in the presence of quarks the 
estimates were significantly different. To improve the convergence for this 
case one could introduce masses for the quarks either by hand or by extending 
the LCO formalism to include the analogous mass operator vacuum expectation 
value $\langle \bar{\psi} \psi \rangle$ which is clearly beyond the scope of 
the present article. Moreover, if one accepts that a gluon mass emerges 
dynamically in QCD, one would then have to include gluon mass corrections in 
the estimates of the pole mass of the quarks. Although we have followed one 
procedure to deduce estimates for the gluon pole mass, other methods are 
possible. Indeed knowledge of the full momentum dependence of the gluon 
$2$-point function would allow for the possibility of repeating the two loop 
analysis which was carried out in \cite{27} using Grunberg's method of 
effective charges, \cite{38,39}. 

\vspace{1cm}
\noindent
{\bf Acknowledgement.} The author thanks R.E. Browne for useful discussions and
Prof. M.Yu. Kalmykov for advice and discussions concerning the {\sc On-Shell 2}
package, \cite{28,29}. 


\end{document}